\documentclass[sigconf, natbib=true, nonacm]{acmart}

\usepackage{booktabs} 
\usepackage{multirow} 
\usepackage{enumitem}
\usepackage{array}
\usepackage{graphicx}

\AtBeginDocument{%
  }

\begin{document}

\title[Exploring Serendipity in Information Seeking for Digital Collections]{Exploring Serendipity in Information Seeking for Digital Collections: A Mixed-Methods Survey Study toward Human-Centered Design}

\author{Saumik Shashwat}
\email{saumiks@icloud.com}
\orcid{0009-0003-0900-8324}
\authornote{These authors contributed to this study equally.}
\affiliation{%
  \department{Human Centered Design and Engineering}
  \institution{University of Washington}
  \city{Seattle}
  \state{WA}
  \country{USA}
}

\author{Xiaoyi Xue}
\email{xiaoyixue0119@gmail.com}
\orcid{0009-0002-1930-471X}
\authornotemark[1]
\affiliation{%
  \department{Human Centered Design and Engineering}
  \institution{University of Washington}
  \city{Seattle}
  \state{WA}
  \country{USA}
}

\author{Benjamin Lee}
\email{bcgl@uw.edu}
\orcid{0000-0002-1677-6386}
\affiliation{%
  \department{Information School}
  \institution{University of Washington}
  \city{Seattle}
  \state{WA}
  \country{USA}
}

\renewcommand{\shortauthors}{Shashwat et al.}

\begin{abstract}
In this paper, we explore user experience and perception of serendipity in information seeking for digital collections through a human-centered design lens. Beginning with an exploratory scoping literature review, we collated the theoretical foundations of serendipity, serendipity in information seeking, search user interfaces, and digital collections. By positioning researchers interacting with digital collections as our primary stakeholders, we utilized a mixed-methods approach involving an online survey study (N = 30). We primarily inquired study participants about the digital collections they worked with, their information seeking behavior, system user experience, and perception of serendipity. Results show that participants with both broad and specific goals, depending on the situation, reported a higher perception of serendipity than participants with specific goals. We found correlations between certain aspects of serendipitous digital environment and perception of serendipity, along with the aspects of former that influence the latter: the system providing more opportunities for unexpected interactions with information, ideas, or resources while seeking information in digital collections corresponded to higher user perceptions of serendipity. We also found correlations between serendipity and autobiographically-identified key research metrics, including a strong positive alignment among learning, collaboration, and value change. Moreover, results indicated mixed sentiments toward AI-facilitated serendipity features. Lastly, possible system design directions to facilitate serendipity in information seeking are discussed. 
\end{abstract}



\maketitle

\section{Introduction}

    Various scholars and previous works have attempted to define and aggregate the definitions for serendipity and its relevance to library \& information science and the digital humanities \cite{de_bruijn_new_2008, andre_x-rays_2009, andre_discovery_2009, liang_designing_2012, agarwal_towards_2015, carr_serendipity_2015, martin_role_2016, mccay-peet_researching_2018, copeland_serendipity_2023, mon_enabling_2024}. However, Kotkov et al. \cite{kotkov_survey_2016} noted a lack of systematic definitions and consensus for serendipity and related concepts, and these shortfalls make it difficult to build theory. Building upon this noted gap, McCay Peet and Toms \cite{mccay-peet_examination_2015} noted, \textit{``inconsistencies exacerbated by the fact that the various disciplines and fields examining serendipity (e.g., information behavior, recommender systems) appear to be working in silos—odd considering the boundary-crossing often required to spark the very phenomenon under investigation.''} We approach this major research gap, especially in the context of information seeking in digital collections, echoing various focal points through human-centered design. Our approach allows for a multidisciplinary epistemological framework to anchor the target users, in this case, the researchers at the core, as we explore the problem space. Our mixed-methods investigations allow us to inquire into their perceptions and experiences. In this work, we first lay down the theoretical foundations connecting serendipity, information seeking, search user interfaces, and digital collections. Following this, we describe our user study methodology, and then report and discuss the results. Our motivations for this paper come from our vision where systems designed to facilitate serendipity in information seeking transform the researcher's experience and nurture key research metrics--\textbf{learning, creativity, collaboration, and value change}-- in research, which are selected based on our autobiographical approach as researchers in the field.

        \begin{figure}[h]
          \centering
          \includegraphics[width=1\linewidth]{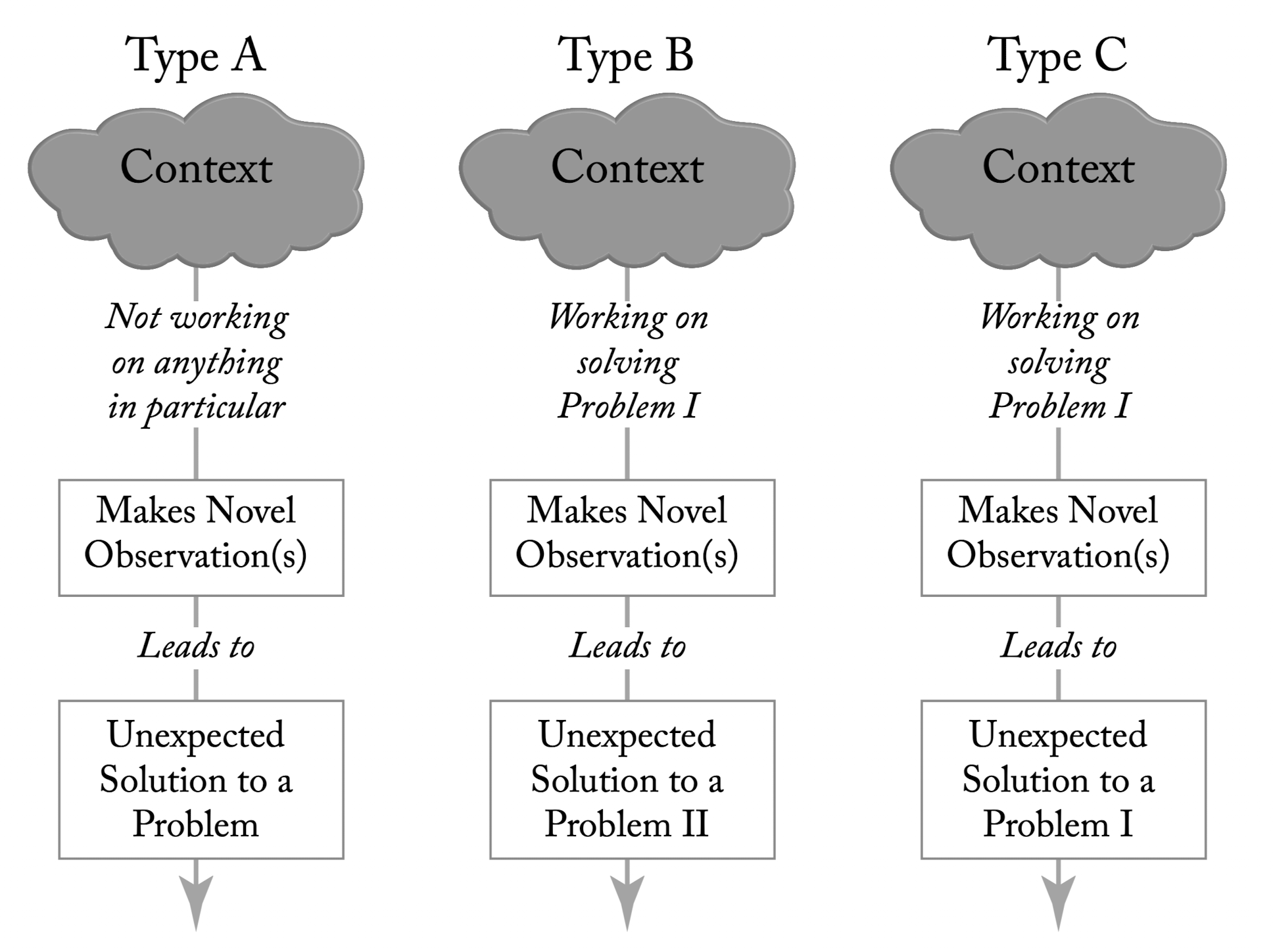}
          \caption{Three ways serendipity happens: types A, B, and C \cite{mccay-peet_examination_2015}.}
          \label{fig:serendipity}
          \Description{Three ways serendipity happens: types A, B, and C. All three types begin with a certain context. In type A, one is not working on anything in particular, makes novel observation(s) which leads to an unexpected solution to the problem. In type B, one is working on solving problem 1, makes novel observation(s) which leads to an unexpected solution to problem 2. In type C, one is working on solving problem 1, makes novel observation(s) which leads to an unexpected solution to problem 1.}
        \end{figure}

\section{Background}
    \subsection{Theoretical Foundations of Serendipity}
        Our theoretical foundation for the study is primarily guided by McCay Peet and Toms' attempt at aggregating existing knowledge in the context of serendipity in digital environments \cite{mccay-peet_examination_2015}. They bifurcate serendipity as primarily either: 1) a quality of an event, something, or someone; or 2) a process or experience, which has one or more serendipitous qualities. The authors noted the former attribution is more prominent in research relating to recommender systems and search engines, while the latter is more holistic and prominent in research relating to information sciences and human-computer interaction. 

        Understanding more about \textbf{event, something, or someone}, the authors present examples of previous research exploring as follows. For serendipity as a quality of an event, previous works have explored the \textbf{`event'} as an event, accident, encounter, interaction, exposure, acquisition, discovery, or conversation. For serendipity as a quality of something, previous works have explored the \textbf{`something'} as recommendations, search results, social networks, and digital environments. For serendipity as a quality of someone, previous works have explored the \textbf{`someone'} as the individual or the information behavior \cite[p. 27, 29, 31]{mccay-peet_examination_2015}.
        
        Understanding more about \textbf{process or experience}, the authors note three main stages that are evident in their selected five serendipity models and frameworks \cite{corneli_modelling_2020, makri_coming_2012, mccaypeet_investigating_2015, rubin_facets_2011, xu_sun_user-centred_2011}: (1) trigger/connection; (2) follow-up; and (3) valuable outcome/reflection. \textit{``The trigger/connection and valuable outcome elements are present in all five models in one form or another. Something or someone must be the catalyst for the experience (i.e., trigger), the user must make a connection between the trigger and their knowledge or experience (i.e., prepared mind), and it must have a valuable outcome because, after all, serendipity is by nature a positive experience.''} \cite[p. 24]{mccay-peet_examination_2015}. It is interesting to note here that all of the aforementioned five models share optimism towards the user experience around serendipity.
    
        In our paper, our approach combines the aforementioned bifurcation as we explore the underlying aspects of serendipity in information seeking in search user interfaces of digital collections.
        
        The authors also describe serendipity to be of three types \cite[p. 9]{mccay-peet_examination_2015}, as shown in Fig. \ref{fig:serendipity}: Type A) When an undefined problem context leads to a novel observation that results in an unexpected solution to a problem; Type B) When the defined problem context for problem 1 leads to a novel observation that results in an unexpected solution to a problem 2; and Type C) When the defined problem context for problem 1 leads to a novel observation that results in an unexpected solution to the same problem, coming from an unexpected source (also referred to as pseudo-serendipity) \cite{mccay-peet_researching_2018}.

        This distinction shapes our understanding of the serendipity types for designing our testing environments and observing the serendipitous phenomena.
    
    \subsection{Serendipity in Information Seeking}
\begin{figure*}[h]
          \centering
          \includegraphics[width=0.2\linewidth]{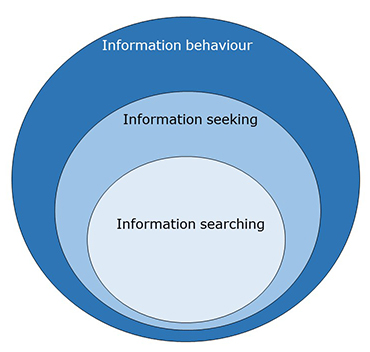}
          \includegraphics[width=0.39\linewidth]{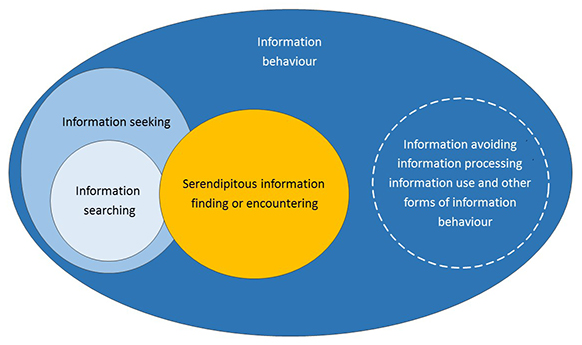}
          \includegraphics[width=0.39\linewidth]{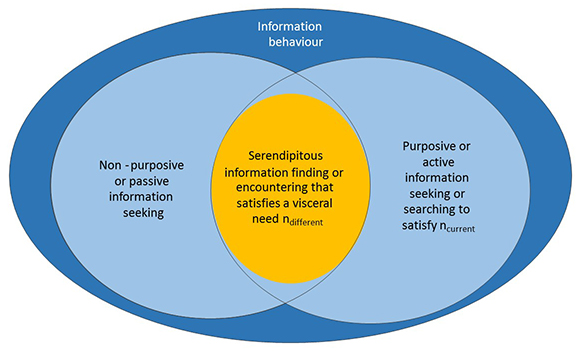}
          \caption{From L to R: A) Wilson's (1999) nested model; B) Placing serendipity within information behavior; C) Placing serendipity within information behavior when not actively seeking and during active seeking or search. \cite{agarwal_towards_2015}}
          \label{fig:agrawal}
          \Description{From L to R: A) Wilson's (1999) nested model where information searching is placed inside information seeking, which is placed inside information behavior; B) Placing serendipity within information behavior; C) Placing serendipity within information behavior when not actively seeking and during active seeking or search.}
        \end{figure*}

        In the paper \textit{Towards a definition of serendipity in information behaviour} \cite{agarwal_towards_2015}, Agarwal utilizes Wilson's model that shows the relationship between information searching, information seeking, and information behavior (1999) as a theoretical lens (Fig. \ref{fig:agrawal}A), to place serendipity within other models of information behavior (Fig. \ref{fig:agrawal}B). Depending on the immediacy of the information need that the encountered information satisfies, the author proposed a nested model for serendipity (Fig. \ref{fig:agrawal}C).

        Agarwal defines serendipity in information behavior as \textit{``...an incident-based, unexpected discovery of information leading to an aha! moment when a naturally alert actor is in a passive, non-purposive state or in an active, purposive state, followed by a period of incubation leading to insight and value.''} However, the author notes approaching this definition with three key assumptions: \textit{``1) The serendipitous discovery itself is always an incident or trigger of a chance encounter, followed by a period of incubation which leads to attaching insight and value to the information encountered. This happens when a visceral or conscious information need is satisfied. 2) Serendipitous information finding leads to unexpectedness, surprise, or an ahah! moment but could lead to disappointment as well. 3) The degree of surprise (ahah! moment) will be higher when one is in a state of natural alertness, as opposed to being in a state of serendipitous alertness or a prepared mind.''} This understanding informs our user study design and the analyses that follow.

    \subsection{Search User Interfaces}
        In the book \textit{Search User Interfaces} \cite{hearst_search_2009}, Marti Hearst notes, \textit{``The job of the search user interface is to aid users in the expression of their information needs, in the formulation of their queries, in the understanding of their search results, and in keeping track of the progress of their information seeking efforts.  However, the typical search interface today is of the form typekeywords-in-entry-form, view-results-in-a-vertical-list.''} This observation made by 2009 still seems to hold true in many search user interfaces we regularly interact with in today's time, especially with digital cultural heritage collections such as library collections, which often rely on controlled vocabularies and subject headings for browsing \cite{library_of_congress_subject_2025}. The author also noted some reasons for the relative simplicity and unchanging nature of the standard Web search interface, attributing it to, \textit{``1) Search is a means towards some other end, rather than a goal in itself.  When a person is looking for information, they are usually engaged in some larger task, and do not want their flow of thought interrupted by an intrusive interface. 2) Related to the first point, search is a mentally intensive task. When a person reads text, they are focused on that task; it is not possible to read and to think about something else at the same time. Thus, the fewer distractions while reading, the more usable the interface. 3) Since nearly everyone who uses the Web uses search, the interface design must be understandable and appealing to a wide variety of users of all ages, cultures, and backgrounds, applied to an enormous variety of information needs.''} These attributions may be experienced and connected differently when we consider search user interfaces in digital collections as opposed to web search interfaces: for example, search can still be a means towards some other end in an interface design that is digital collection-centered. This design can still be understandable and appealing but maybe majorly for a user group with higher expertise and familiarity. We believe that in doing so, the rather non-simplistic and changing (to pertain to the nature of the collection) design of a search user interface itself can facilitate serendipity, which can lead to incident-based, unexpected discoveries of information.
    
    \subsection{Digital Collections}
        Within the context of cultural heritage, digital collections are an essential form of access for information seekers, ranging from academic researchers to journalists, from teachers and students in the classroom to curious members of the public. These collections--comprising digitized and increasingly born-digital materials--are made available and discoverable by institutions including galleries, libraries, archives, and museums. Despite consistently active user bases, however, digital heritage collections are often overlooked within the information seeking literature in favor of other settings. We contend that adopting digital  collections as our setting for serendipity within information seeking therefore represents an important direction for research within human-computer interaction and exploratory search, with findings that may be relevant in informing our understanding of serendipity when browsing more generally.

\section{Methodology}
    We conducted a mixed-methods survey study to understand the underlying aspects of serendipity in information seeking in search user interfaces of digital collections. The survey structure was divided into: 1) digital collections, 2) information seeking, and 3) serendipity. McCay Peet and Toms \cite[p. 48]{mccay-peet_researching_2018} laid down the purpose-method mappings for measuring various facets of serendipity with certain methods: our study fulfills 4 out of 9 purposes (2-4, 9). In this paper, we refer to the survey respondents as `participants', and the search user interface of a digital collection as the `system'. In the following subsections, we elaborate on participant demographics, recruitment process, survey questions, measurements, data analyses, ethics, and consent.

    \subsection{Participants and Recruitment}
        The survey was hosted on Google Forms and collected responses from September 2025 to January 2026. We recruited a total of 30 participants (ages ranging from 19 to 68) through convenience sampling \cite{etikan_comparison_2015} majorly via university channels, listservs, and social media. We also collected participants' demographic information, including the highest completed education level, domains of expertise, profession, ethnic origins or ancestry, and language(s) they can read and/or speak other than English, and recruitment for future studies in this survey. These two segments were optional and did not affect their participation in the study.

    \subsection{Questions and Measurements}
        For the survey, we defined: 1) a \textbf{digital collection} as any digital (born digital or digitized) collection served by a gallery, library, archive, or museum; 2) \textbf{information seeking} as a conscious effort to acquire information in response to a need, want, or gap in our knowledge \cite{case_looking_2012}; and 3) \textbf{information searching} as \textit{``a subset of information seeking, particularly concerned with the interactions between information user [actor]… and computer-based information systems''} \cite[p. 263]{wilson_models_1999}. In what follows, we further describe the protocol, structure, and reasoning. The full survey questionnaire can be found in the Appendix, Section \ref{sec:questionnaire}.

        \subsubsection{Digital Collections}
            Our survey was centered on the participant-reported most frequented digital collection (DC). We also inquired if the reported DC was `Born-digital', `Digitized', or `Both'.

        \subsubsection{Information Seeking}
            We asked participants about their primary reasons for seeking information in the reported DC, henceforth referred to as \textbf{information seeking goals}. We also asked their \textbf{information searching orientation} when they begin an information search in the reported DC: do they have a `Specific goal' in mind, do they perform a `Broad exploration', or `Both, depending on the situation'? Another item inquired about the specific \textbf{information searching strategies} participants typically use when searching in the reported DC. 
            
            There were two additional qualitative items asking about the system user experience with the pain points in information seeking and suggestions in improving this overall experience. 
            
        \subsubsection{Serendipity}
            To understand and quantitatively evaluate the occurrence of serendipity and its perception in the user-reported digital collections, our survey incorporated the \textbf{Serendipitous Digital Environment (SDE)} and \textbf{Perception of Serendipity (PoS)} scales. Both scales were measured on a 5-point Likert scale of agreement (1: Strongly disagree, 2: Disagree, 3: Neither agree nor disagree, 4: Agree, 5: Strongly agree) \cite{mccay-peet_examination_2015, mccay-peet_researching_2018}. For the purpose of our evaluation, the individual nominal items were grouped at factor and scale levels to latent construct composite scale values.
    
            Our survey questionnaire for the former scale consisted of 15 items in total, which were categorized into four SDE factors: `Leads to the Unexpected' (LU), `Trigger-Rich' (TR), `Highlights Triggers' (HT), and `Enables Connections' (EC) \cite{mccay-peet_examination_2015}. These factors compute the user's assessment of the degree to which their reported digital environment:
            \begin{enumerate}
                \item provides opportunities for unexpected interactions with information, ideas, or resources: LU (5 items);
                \item contains a variety of information, ideas, or resources that is interesting and useful to the user: TR (3 items);
                \item brings interesting and useful information, ideas, or resources to the user's attention: HT (4 items);
                \item makes relationships or connections between information, ideas, or resources apparent: EC (3 items).
            \end{enumerate}
    
            Our questionnaire for the latter scale consisted of 4 items inquiring into the `specific digital environment' \cite{mccay-peet_examination_2015} users reported earlier in the survey:
            \begin{enumerate}
                \item In (this digital collection), I experience serendipity that has an impact on my everyday life;
                \item (...), I experience serendipity that has an impact on my work;
                \item I encounter useful information, ideas, or resources that I am not looking for when I use (...);
                \item In (...), I experience mixes of unexpectedness and insight that lead to valuable, unanticipated outcomes.
            \end{enumerate}

            There were two additional qualitative questions about AI-facilitated serendipity features in the reported DC and additional thoughts about serendipity and DC that participants would like to share. We reiterate that the full survey questionnaire can be found in the Appendix, Section \ref{sec:questionnaire}.

    \subsection{Data Analyses}

        Our survey comprised both quantitative and qualitative inquiries. Hence, we utilized a mixed-methods approach to analyze the study data. For its quantitative counterpart, the authors of this paper performed descriptive and inferential data analyses using IBM SPSS. For qualitative data, one author of this paper performed thematic analyses. After individually performed open coding for each corresponding question, the author shared the results and have them consolidated with other authors.
    
    \subsection{Ethics and Consent}

        All participants voluntarily filled out the study survey and were provided informed consent prior to their participation. Participants were informed that their responses would be used solely for research purposes in accordance with institutional ethical guidelines. Participants have been allowed to refuse to participate and withdraw from this study at any time without penalty or loss of benefits to which they are otherwise entitled. All these recorded materials were stored only for the study team, and any results derived from them have been anonymized in publishing. No cloud services or artificial intelligence services were used in the analyses of participants' data. This study was deemed IRB exempt by the University of Washington Human Subjects Division.

\section{Results and Discussion}
    \subsection{Digital Collections}
        Table \ref{tab:DC} shows all reported digital collections ($N=30$), its type, and participants' information searching orientation. Out of all reported collections, 6 collections ($20\%$) are born-digital collections, 16 ($53.3\%$) are digitized, while 8 ($26.7\%$) are both. Though many of these collections are available online, three respondents included personal collections as their collections of interest.

        \begin{table*}[h]
            \centering
            \caption{Participant-reported Digital Collections}
            \label{tab:DC}
            \begin{tabular}{llll}
                \toprule
                PID & Digital Collection & DC Type & Information Searching Orientation \\
                \midrule
                P1  & British Newspaper Archive \cite{noauthor_british_2026} & Digitized & Specific goal \\
                P2  & The National Archives \cite{archives_national_2026} & Digitized & Specific goal \\
                P3  & National Folklore Collection \cite{noauthor_duchasie_2026} & Digitized & Both, depending on the situation \\
                P4  & Library and Archives Canada \cite{canada_library_2025} & Both & Both, depending on the situation \\
                P5  & NASA Scientific Visualization Studio \cite{studio_nasa_2000} & Born digital & Both, depending on the situation \\
                P6  & Hemeroteca Digital \cite{noauthor_bndigital_2026} & Digitized & Specific goal \\
                P7  & Chronicling America Historic American Newspapers \cite{noauthor_chronicling_2026} & Digitized & Both, depending on the situation \\
                P8  & Chronicling America Historic American Newspapers \cite{noauthor_chronicling_2026} & Digitized & Specific goal \\
                P9 & International Tracing Service Digital Archive \cite{noauthor_arolsen_2026} & Digitized & Specific goal \\
                P10 & USHMM Digital Archives \cite{noauthor_collections_2026-1} & Digitized & Specific goal \\
                P11 & The Newberry's Edward E. Ayer Collection \cite{noauthor_american_2026} & Both & Both, depending on the situation \\
                P12 & Women Writers Online \cite{project_women_2026} & Digitized & Both, depending on the situation \\
                P13 & University of North Texas Scholarly Works Collection \cite{noauthor_unt_2026} & Both & Specific goal \\
                P14 & USHMM Archival Holdings of AJ110 (RG-49.005M) \cite{noauthor_collections_2026} & Digitized & Specific goal \\
                P15 & Civil War Photo Sleuth \cite{noauthor_civil_2026} & Digitized & Both, depending on the situation \\
                P16 & Outskirts of the City (Personal Collection) \cite{noauthor_outskirts_2026} & Digitized & Specific goal \\
                P17 & Notion Workspace (Personal Collection) & Born digital & Specific goal \\
                P18 & Game Arts in League of Legends & Born digital & Specific goal \\
                P19 & Project Research Materials (Personal Collection) & Both & Both, depending on the situation \\
                P20 & Wikipedia & Born digital & Specific goal \\
                P21 & The Emilio Segrè Visual Archives \cite{noauthor_emilio_2026} & Both & Both, depending on the situation \\
                P22 & Library of Congress Prints \& Photographs Online Catalog \cite{noauthor_prints_2026} & Digitized & Both, depending on the situation \\
                P23 & Library of Congress Digital Collections \cite{noauthor_digital_2026} & Both & Both, depending on the situation \\
                P24 & Lessing J. Rosenwald Collection \cite{noauthor_lessing_2026} & Digitized & Broad exploration \\
                P25 & ACM Digital Library \cite{noauthor_acm_2026} & Born digital & Both, depending on the situation \\
                P26 & Chronicling America Historic American Newspapers \cite{noauthor_chronicling_2026} & Digitized & Specific goal \\
                P27 & Washington State Archives - Digital Archives \cite{noauthor_washington_2026} & Born digital & Specific goal \\
                P28 & EBSCO Database \cite{noauthor_ebsco_2026} & Digitized & Both, depending on the situation \\
                P29 & University of Washington Libraries \cite{noauthor_uw_2026} & Both & Specific goal \\
                P30 & Spotify \cite{noauthor_spotify_2026} & Both & Both, depending on the situation \\
                \bottomrule
            \end{tabular}
        \end{table*}

    \subsection{Information Seeking}
        \subsubsection{Reasons for seeking information} Participants reported `research for academic or professional projects' ($n=26; = 86.67\%$ responses), `personal learning and curiosity' ($n=13; = 43.33\%$ responses), `sharing with others' ($n=10; = 33.33\%$ responses), and `making creative decisions' ($n=8; = 26.67\%$ responses) as the top four reasons they seek information in their reported digital collections.

        \subsubsection{Information Searching Orientation} When seeking information inside their reported digital collection (DC), $n=15$ participants ($50\%$ responses) reported having a `specific goal' (SG) in mind, $n=1$ ($= 3.33\%$ responses) reported having a `broad exploration' (BE) goal, while $n=14$ ($= 46.67\%$ responses) reported `both, depending on the situation' (BDS). Henceforth, we will present insights between the SG and BDS groups of responses.

        \subsubsection{Information Searching Strategies} Participants reported often using a combination of strategies; the top five strategies they reported are `keyword searching' ($n=23; = 76.67\%$ responses), `using filters or advanced search options' ($n=21; 70\%$ responses), `browsing categories or tags' ($n=16; = 53.33\%$ responses), following `links or related resources' ($n=15; 50\%$), and lastly, `searching visually' ($n=13; = 43.33\%$ responses)\footnote{The high usage of faceted search methods confirms a wealth of findings about the value of faceted search \cite{yee_faceted_2003}.}.

        \subsubsection{Pain Points in Information Seeking and Suggestions for Improving the Experience} In the responses about pain points in information seeking in DC and suggestions for improving the experiences, two out of sixty combined responses mentioned serendipity (P6 and P12). P6 said, \textit{``I like that I can browse neighboring objects by call number and I can also search by medium, creator, and subject.''} P12 mentioned, \textit{``Honestly, I hope the library begins using tagging a bit more to connect related materials.''} This result showed that serendipity was rarely mentioned in both pain points and suggestions for improving the collection.
    
    \subsection{Serendipity}
        \begin{table}
            \centering
            \caption{Reliability and Descriptive Approximates ($N = 30$)}
            \label{tab:overalldesc}
            \begin{tabular}{llll}
                \toprule
                Scale-\textit{subscale} & $\alpha$ & Mean & SD \\
                \midrule
                SDE & .80 & 3.72 & 0.50 \\
                SDE-\textit{LU} & .88 & 3.77 & .80 \\
                SDE-\textit{TR} & .86 & 4.48 & .63 \\
                SDE-\textit{HT} & .77 & 3.26 & .83 \\
                SDE-\textit{EC} & .89 & 3.51 & 1.00 \\
                PoS & .75 & 3.68 & .75 \\
                \bottomrule
            \end{tabular}
        \end{table}
        
        Overall ($N=30$), Cronbach’s alpha indicated acceptable to good reliability for all our composite `Serendipitous Digital Environment' (SDE) subscales and `Perception of Serendipity' (PoS) scale, as shown in Table \ref{tab:overalldesc}. All noted mean scores depict an above-average to positive assessment of the occurrence and perception of serendipity across the pool of participant-reported digital collections. 

        Shapiro–Wilk tests indicated that `Leads to the Unexpected' (LU) ($p=.13$) and `Highlights Triggers' (HT) ($p=.69$) did not significantly deviate from normality, while `Trigger-Rich' (TR) ($p<0.001$), `Enables Connections' (EC) ($p=.023$), and PoS ($p=.033$) showed significant deviations from a normal distribution.

        \subsubsection{Effect of Information Searching Orientation on User Experience and Perception of Serendipity}
            In our between-group comparison between SG ($n=15$) and BDS ($n=14$), we used independent two-sample t-tests for parametric subscales LU and HT, and Mann-Whitney tests for non-parametric scales TR, EC, and PoS. Only PoS showed a significant group difference ($U = 48.5$, $Z = -2.49$, $p = .013$, $r = .46$), with the BDS group reporting higher scores. This shows that participants with both broad exploration and specific goals, depending on the situation, while searching information in digital collections, reported a higher perception of serendipity than participants with specific goals. We discuss the specific aspects of the system design that may have contributed to this finding in the following subsubsection. 

        \subsubsection{Correlations between Aspects of Serendipitous Digital Environment and Perception of Serendipity}
            We conducted Spearman correlations to examine associations among the SDE subscales and PoS scale. Results revealed that LU was strongly and positively correlated with PoS ($\rho = .72$, $p < .001$). HT showed a moderate positive correlation with EC ($\rho = .38$, $p = .044$). No other correlations reached statistical significance ($p > .05$). This shows that the system providing more opportunities for unexpected interactions with information, ideas, or resources while seeking information in digital collections corresponds to higher user perceptions of serendipity. Additionally, the system bringing forward useful information to users' attention potentially corresponds to the system making relationships or connections between information, ideas, or resources apparent. 
        
        \subsubsection{Aspects of Serendipitous Digital Environment Influencing Perception of Serendipity}
            We also conducted a multiple regression to examine whether the SDE subscales can predict PoS. The model was significant ($F(4, 24) = 9.21$, $p < .001$), explaining $60.6\%$ of the variance in PoS (adjusted $R ^ 2 = .54$). LU emerged as a strong positive predictor of PoS ($B = .64$, $\beta = .68$, $p < .001$), and EC emerged as a significant positive predictor ($B = .23$, $\beta = .31$, $p = .044$). TR and HT were statistically insignificant predictors ($p > .50$). This might indicate that participants' perception of serendipity while seeking information in digital collections is primarily driven by the system's ability to provide opportunities for unexpected interactions with information, ideas, or resources, and to make connections between them apparent to the user.

        \subsubsection{Serendipity and Key Research Metrics}
            We conducted ordinal logistic regression to examine the relationship between information seeking orientation (SG vs. BDS) and researchers' self-reported association between serendipity and four key research metrics (Likert scale 1-5, ranging from Strongly Disagree to Strongly Agree), namely \textbf{learning, creativity, collaboration, and value change}. Multinomial logistic regression was used for learning because the proportional odds assumption was violated in ordinal logistic regression ($p = .042 < 0.05$). Results revealed no correlation between information searching orientation and Learning ($x^2 = 4.164$, $p = .125 > .05$), Creativity ($x^2 = 0.782$, $p = .376 > .05$), Collaboration ($x^2 = 1.165$, $p = .558 > .05$), and Value Change ($x^2 = 1.493$, $p = .222 > .05$). This indicated there is no correlation between the type of information seeking that researchers choose to conduct and participants' self-reported association between serendipity and key research metrics. 
        
            We conducted Spearman’s rank-order correlation within researchers' self-reported association between serendipity and four key research metrics to examine the potential correlations among the metrics. Results indicated: (1) a strong positive association between learning and value change ($\rho = .718$, $p < .001$, $n=29$), which is the highest among the associations; (2) a strong positive association between collaboration and value change ($\rho = .679$, $p < .001$, $n=29$); (3) a strong positive association between learning and collaboration ($\rho = .610$, $p < .001$, $n=29$). This shows that in participants' self-reported association between serendipity and key research metrics, there is a strong positive alignment among learning, collaboration, and value change. Creativity, however, does not share a similar trend with the other three metrics (for learning: $\rho = .373$, $p = .046 > .001$, $n=29$; for collaboration: $\rho = .386$, $p = .038 > .001$, $n=29$; for value change: $\rho = .154$, $p = .425 > .001$, $n=29$). This finding is intriguing, and we will explore more in our future studies.  

        \subsubsection{AI-facilitated Serendipity}
            To explore AI-facilitated serendipity features in reported DC, we asked the following question in the form of and open-ended question: \textit{``Have you ever come across or used any AI-facilitated serendipity or similar implementations in the digital collection(s) you frequent, or in your information-seeking processes? If so, could you describe them?''}

            In the responses to this question, 18 participants indicated they have not seen AI-facilitated serendipity features in reported DC, while 12 of them indicated they have been exposed to this kind of feature, with 2 of the participants who are not sure. From the responses for those who have not been exposed to AI-facilitated features, major emerging themes are \textbf{General negative attitude towards AI}, and \textbf{AI can be used in DC}. In the thematic analysis of responses for those who have been exposed to AI-facilitated features, emerging themes are \textbf{AI is not helpful in forming serendipity}, \textbf{AI is helpful in forming serendipity type A and B}, and \textbf{AI is helpful in other parts of the collection}. This initial analysis showed whether being exposed to AI-facilitated features does not impact researchers' attitude (positive and negative) towards AI. 

            In the further categorization of emerging themes based on participants' attitude towards AI (positive and negative), \textbf{AI is helpful in forming serendipity type A and B}, \textbf{AI is helpful in other parts of the collection} and \textbf{AI can be used in DC} are mapped to positive. Under the theme of \textbf{AI is helpful in forming serendipity type A and B}, participants expressed that AI helps them form new unknown connections (P15) and unknown websites (P22), which corresponds to the definition of serendipity type one. P22 said, \textit{``Sometimes this is really surprising and gets me to images that are actually a better fit or things that I didn't even realize I wanted to see. And it also gets me to webpages that I wasn't aware of and don't come up when I am doing a word-based search on the same topic. In general, this has been really useful for me.''} For P22, they did not even realize they wanted to see the suggested contents, indicating that they did not have an aim when conducting information seeking. Therefore, it corresponds to serendipity type A. For serendipity type B, where participants have an intention when conducting information seeking, participants mentioned about AI helping in forming new connections from past materials (P15), suggesting useful images (P22) and offering similar visual contents (P22 and P26). P26 used \textit{Newspaper Navigator} \cite{lee_newspaper_2020} as an example and said, \textit{``(it) enabled users to find visually similar content in newspapers using AI. It enabled me to explore and find images I otherwise would not have been able to find.''} Different from P22, P26 described a situation where they are looking for specific visual contents and AI-facilitated serendipity helped them to achieve this goal by suggesting visually similar content, which corresponds to serendipity type B. 

            Additionally to serendipity, participants mentioned AI helped them in understanding complex descriptions (P5), refining searching keywords (P18), organizing materials (P19), and filling in forgotten ideas (P19), under the theme of \textbf{AI is helpful in other parts of the collection}. Moreover, participants expressed their positive attitude towards AI under the theme of \textbf{AI can be used in DC}. They talked about AI can be used in the back scene (P12) and positively impact creativity (P23). 

            Turning to negative attitudes towards AI, \textbf{AI is not helpful in forming serendipity} and \textbf{General negative attitude towards AI} are the major two themes. In the first theme, participants expressed serendipity can only be formed by themselves (P1) and negative serendipity experience when encountering incorrect information produced by AI (P11). P1 said, \textit{``I've encountered other collections that try to build `serendipity' or `connection-making' into their architecture, but I haven't found them helpful at all to be honest. Sometimes they are distinctly unhelpful as they push me toward useless material.''} In terms of general attitude towards AI, participants expressed their ethical concerns (P3) and negative emotions (P28) in incorporating AI in DC. P3 said \textit{``These (digital) collections have not been used for AI because of ethical concerns for the cultural heritage material.''} These sentiments reflect an increasingly large contingent of practitioners in cultural heritage institutions who support forms of AI refusal \cite{fox_ai_2025}.
            
        \subsubsection{Additional Thoughts}
            In our open-ended question to gather additional thoughts about serendipity and DC(s) that participants would like to share, three emerging themes emerged. The major theme was participants expressing \textbf{Serendipity as an insightful research}. P24 described their experiences in details: \textit{``I do not consider myself a scholar or researcher, but have always approached libraries and archives as a curious person looking for new ideas by serendipitously discovering a a title on a book shelf or a dried flower in an archival folder of someone's personal papers, analog moments of magic like that. I like browsing antique stores for the same reason! I'm convinced that there are ways to craft these types of experiences digitally as well, and they are necessary to accessibility and even the utility of digital collections presented online which are often presented as such a confusingly siloed, decontextualized representation of the original accession.''} Participants also expressed \textbf{The relationship between serendipity and DC design}. They talked about serendipity being hard to separate from good design (P8) and providing the concept of generative interfaces as an example (P12). Moreover, participants offered insights in \textbf{Other design ideas for interface}, which includes visualizing physical characteristics (P6) and developing easy ways to browse (P5). These results show the importance of increasing awareness of serendipity in the context of DC. Some of the researchers (n=4) are interested in the topic of serendipity, but not everyone. Also, these results combined with the results of participants' pain points show that participants do not consider the lack of serendipity in DC as a pain point but a \textit{``bonus''} feature to have in the design of DC.

\section{Conclusion}
    In this study, we adopted a human-centered lens to exploring serendipity in information seeking for digital collections. We conducted an online survey ($N = 30$) comprising mixed-methods. For digital collections, we asked participants to report one digital collection, its type, and their searching orientation. For information seeking, we asked questions about reasons for seeking information, information search orientation, information searching strategies, and pain points in information seeking and suggestions for improving the experience. Through questions about serendipity, we used Serendipitous Digital Environment and Perception of Serendipity scales for participants to evaluate their selected collections, and report their self-rated association between serendipity and four key research metrics (learning, creativity, collaboration, and value change). We also covered open-ended questions about AI-facilitated serendipity features and additional thoughts. 

    Results for digital collections showed the overall variety of collections participants used in their research. For information seeking, results showed the multiple reasons participants were seeking information in the system, an almost even spread between `Specific Goal' and `Both, depending on the situation' subgroups, top information searching strategies echoing the value of faceted search, and serendipity not being mentioned as a pain point nor as a suggestion for improving the collection. Results for serendipity show that participants with both broad and specific goals, depending on the situation, reported a higher perception of serendipity than participants with specific goals. We found correlations between certain aspects of serendipitous digital environment and perception of serendipity, along with the aspects of former that influence the latter--the system providing more opportunities for unexpected interactions with information, ideas, or resources while seeking information in digital collections corresponded to higher user perceptions of serendipity. We also found correlations between serendipity and autobiographically-identified key research metrics--learning, creativity, collaboration, and value change--there was a strong positive alignment among learning, collaboration, and value change. Results also explored the nuances of user experiences with AI-facilitated serendipity features. Additionally, participants expressed serendipity as insightful research, its relation with the design of digital collections, and suggested other design ideas for the interface. We discussed possible directions that can be considered for those systems designed to facilitate serendipity in information seeking.

\section{Future Work}
    For next steps, we plan to conduct three follow-on studies: (1) targeted interviews, (2) system walkthroughs, and (3) a co-design workshop. This combination of studies will ensure a full coverage of the nine purpose-method mappings for measuring various facets of serendipity with certain methods \cite[48]{mccay-peet_researching_2018}.
    
    For (1) and (2), we have already conducted eight sessions with a subset of survey respondents to provide more depth and granularity to complement our survey-based approach in this paper. We believe that the richness of the data we have already collected could be valuable to understanding how serendipity emerges within specific search experiences for digital collections. We plan to conduct more interviews and walkthroughs and synthesize our findings. 
    
    After we complete this step, we will conduct a co-design study with hands-on sketching and prototyping of new interfaces and affordances for supporting serendipity within digital collections.
    
    Lastly, in terms of the association between serendipity and key research metrics, we will explore more in our future study. Moreover, there lies value in correlating the demographic insights we collected with our survey and the rest of the study results to inform future research for specific user subgroups.

\bibliographystyle{ACM-Reference-Format}
\bibliography{references}

\appendix
\section{Survey questionnaire: main items}\label{sec:questionnaire}

\begin{enumerate}
    \item List one digital collection you have the most experience working with. (Answer the following questions accordingly)

    \item Is this digital collection a born digital or a digitized collection (existing as physical first; converted to digital later)?
    \begin{itemize}
        \item Born digital
        \item Digitized
        \item Both
        \item Other: \_\_\_\_
    \end{itemize}

    \item What is your level of familiarity with this digital collection (content, structure, etc.), on a scale of 0 to 5 (0: Not familiar, 5: Very familiar)?

    \item How often do you use this digital collection for seeking information?
    \begin{itemize}
        \item Daily
        \item Once per week
        \item Once per month
        \item Once per quarter
        \item Yearly
        \item Never
        \item Other: \_\_\_\_
    \end{itemize}

    \item What are your primary reasons for seeking information in this digital collection? (Select all that apply)
    \begin{itemize}
        \item Research for academic / professional projects
        \item Personal learning and curiosity
        \item Making creative decisions
        \item Sharing with others
        \item Other: \_\_\_\_
    \end{itemize}

    \item When you begin an information search in this digital collection, do you typically have a specific goal in mind, or do you explore more broadly?
    \begin{itemize}
        \item Specific goal
        \item Broad exploration
        \item Both, depending on the situation
        \item Other: \_\_\_\_
    \end{itemize}

    \item Which of the following strategies do you typically use when searching in this digital collection? (Select all that apply)
    \begin{itemize}
        \item Keyword searching
        \item Using filters / advanced search options
        \item Searching visually
        \item Browsing categories / tags
        \item Following links / related resources
        \item Consulting recommendations or suggestions
        \item Other: \_\_\_\_
    \end{itemize}

    \item How often do you encounter pain points / difficulties seeking information in this digital collection?
    \begin{itemize}
        \item Never
        \item Occasionally
        \item Sometimes
        \item Often
        \item Always
        \item Other: \_\_\_\_
    \end{itemize}

    \item If you encounter pain points / difficulties, what are the most common challenges?

    \item Do you have any suggestions for improving the digital collection(s) you work with or information seeking experience in general?

    \item The following 15 items inquire about serendipity in digital environments. Fill in correspondence to the digital collection you have most experience working with, on a scale of 1 to 5 (1: Strongly disagree, 2: Disagree, 3: Neither agree nor disagree, 4: Agree, 5: Strongly agree).

        \begin{enumerate}
            \item I bump into unexpected content in this digital collection.
            \item I come across content by chance in this digital collection.
            \item I am exposed to unanticipated content in this digital collection.
            \item I stumble upon information in this digital collection.
            \item I encounter the unexpected in this digital collection.
            \item This digital collection is full of information useful to me.
            \item I find information of value to me in this digital collection.
            \item This digital collection is a treasure trove of information.
            \item This digital collection has features that draw my attention to information.
            \item This digital collection has features that alert me to information.
            \item This digital collection has features that ensure that my attention is drawn to useful information.
            \item I am pointed toward content in this digital collection.
            \item I can see connections between topics in this digital collection.
            \item This digital collection enables me to make connections between ideas.
            \item I come to understand relationships between ideas in this digital collection.
        \end{enumerate}

    \item The following 4 items inquire about your perception of serendipity. Fill in correspondence to the digital collection you have most experience working with, on a scale of 1 to 5 (1: Strongly disagree, 2: Disagree, 3: Neither agree nor disagree, 4: Agree, 5: Strongly agree).

        \begin{enumerate}
            \item I encounter useful information, ideas, or resources that I am not looking for when I use this digital collection.
            \item In this digital collection, I experience mixes of unexpectedness and insight that lead to valuable, unanticipated outcomes.
            \item In this digital collection, I experience serendipity that has an impact on my everyday life.
            \item In this digital collection, I experience serendipity that has an impact on my work.
        \end{enumerate}

    \item Based on your personal experiences and understanding, how would you describe your associations between serendipity and the following metrics, on a scale of 1 to 5 (1: Strongly negative, 2: Negative, 3: Neutral, 4: Positive, 5: Strongly positive)? (thinking prompt: \textit{``Serendipity is associated with \_\_ for me''})

        \begin{enumerate}
            \item Learning
            \item Creativity
            \item Collaboration
            \item Value Change
        \end{enumerate}

    \item Have you ever come across or used any AI-facilitated serendipity or similar implementations in the digital collection(s) you frequent, or in your information-seeking processes? If so, could you describe them?

    \item Do you have any additional thoughts about serendipity and digital collection(s) you would like to share?

\end{enumerate}

\end{document}